\documentclass[fleqn,10pt]{wlscirep}
\title{Signatures of Parkinson's disease in complexity of low-frequency fluctuations of postural  sway velocity}

\author[1,+]{Miroslaw Latka}
\author[2,+*]{Slawomir Budrewicz}
\author[1,+]{Klaudia Kozlowska}
\author[2,+]{Magdalena Koszewicz}
\author[3,+]{Bruce J. West}

\affil[1]{Department of Biomedical Engineering,  Faculty of Fundamental Problems of Technology, Wroclaw University of Science and Technology, Wroclaw, 50-370, Poland.}
\affil[2]{Department of Neurology,Wroclaw Medical University, Borowska 213, 50-556 Wroclaw, Poland.
}
\affil[3]{Army Research Office, Information Sciences Directorate, Research Triangle Park, 27709, USA.}

\affil[*]{Miroslaw.Latka@pwr.edu.pl}

\affil[+]{each author contributed equally to this work}

\keywords{Parkinson's disease; posturography; Lempel-Ziv complexity}

\begin{abstract}

Posturography is routinely used to qualitatively assess one of the cardinal symptoms of Parkinson's disease  -- postural instability. While most measures of balance control are derived from displacement of the center of pressure there is evidence that such control is more likely to be velocity-based. We performed static posturographic tests (eyes open and eyes closed) during quiet standing in narrow stance for \textit{n}=30 patients with Parkinson's disease (PDP) in the ON state and compared the results with those of \textit{n}=30 age-matched senior controls (HSC) and \textit{n}=60 young controls (HYC). We used differentiator filters to generate time series of low-frequency fluctuations of sway velocity and calculated their Lempel-Ziv complexity (LZC). With eyes closed, the mediolateral LZC of HSC  0.21 (0.02) was significantly higher than those of HYC 0.19 (0.02) and PDP 0.18 (002). Thus aging and PD have \textit{opposite} effects on mediolateral LZC which strongly differentiates between HSC and PDP (92\% sensitivity and 87\% specificity).

\end{abstract}

\begin{document}

\flushbottom
\maketitle

\thispagestyle{empty}

\section*{Introduction}

2017 marks the 200th anniversary of James Parkinson's \textit{Essay on the Shaking Palsy}. In the intervening years remarkable progress has been made in the characterization of the phenomenology of Parkinson's disease (PD), the elucidation of underlying neurodegenerative mechanisms and the development of medical and surgical treatments. Despite these advances, it is difficult to overestimate the burden imposed by this disease on patients, their families and society at large. PD is the second most common neurodegenerative disorder, after Alzheimer's disease, with an overall prevalence of 300 cases per 100,000 \cite{Kalia2015}.

The prominent death of dopaminergic neurons in the substantia nigra pars compacta (SNpc) is a defining feature of PD. The resultant dopamine deficiency within the basal ganglia leads directly to bradykinesia, muscular rigidity and rest tremor. The other motor symptoms: impairment of posture, balance and gait, are largely secondary to degeneration of nondopaminergic pathways \cite{Magrinelli2016}. Posturography is frequently used to qualitatively assess postural instability. A number of measures of balance control are derived from spontaneous fluctuations of the center of pressure (COP) during quiet standing. While the planar COP trajectory does reflect some aspects of an underlying control process, both theoretical analysis and experiments indicate that balance control is more likely to be velocity-based than position-based \cite{Kiemel2002,Masani2003, Jeka2004}. Leaving this controversy aside, mean sway velocity calculated, for example, as the ratio of the COP path length and the duration of the measurement, has long been recognized not only as the most reliable among traditional measures of balance control \cite{Cornilleau-Peres2005,Raymakers2005,Lin2008}, but also as a sensitive marker of developmental changes \cite{Riach1987,Riach1994,Kirshenbaum2001,Rival2005,Chen2008,Ajrezo2013} and aging \cite{Baloh1994,Prieto1996,Vseteckova2013}. With the exception of the study of Delignieres et al. \cite{Delignieres2011} spontaneous fluctuations of sway velocity have not been investigated. This is surprising since the statistical properties of physiological fluctuations, like those found in the time series for heartbeat dynamics, respiration, human locomotion and posture control, have been the focus of interdisciplinary research for more than two decades. The loss of physiological complexity has been observed in various diseases and in aging \cite{Goldberger, Vaillancourt2002,Manor2013}. Deterministic (and often periodic) temporal dynamics is one possible realization of loss of physiological complexity \cite{goldberger1990chaos}. Maurer et al. observed 1 Hz body sway oscillations in PD patients in the OFF state. These oscillations vanished under levodopa treatment and/or deep brain stimulation \cite{Maurer2003,Maurer2004}.

The goal of the present study is to determine the influence of aging and PD on the complexity of low-frequency (smaller than $\sim $1.5 Hz) sway velocity fluctuations, during quiet standing. We hypothesize that a decrease in complexity is a distinct feature of PD pathophysiology, which can be detected even in the ON state. This reduction in complexity is consistent with an early prediction of the relation between physiologic complexity and disease \cite{goldberger1990chaos}.

\section*{Results}

In Table \ref{tab:statistics} we collected the values of low-frequency speed $|v|$, Lempel-Ziv complexity ($LZC$, equation \ref{LZC}) of low-frequency velocity $v$ as well as  body sway velocity ($BSV$) for mediolateral (\textit{x}) and anteroposterior (\textit{y}) directions. The values were averaged over three cohorts: healthy senior controls (HSC), healthy young controls (HYC) and patients with Parkinson's disease (PDP). For each of these three quantities we calculated  the indices which quantify the contribution of vision to balance control: the Romberg ratio $R$ (equation \ref{R}) and the stabilization ratio $SR$ (equation \ref{SR}). In Table \ref{tab:significance} plus signs indicate statistically significant differences between cohorts (cases for which the posthoc analysis yielded values $p<0.05$). The data presented in Tables \ref{tab:statistics} and \ref{tab:significance} are crucial for testing the main hypothesis of the paper: that a decrease in complexity of low-frequency fluctuations of postural sway velocity is a signature of Parkinson's disease.

\begin{table}
   \centering
   \caption{Group averaged values of speed $|v|$,  Lempel-Ziv complexity \textit{LZC} of low-frequency velocity  and body sway velocity $BSV$ for mediolateral (\textit{x})  and anteroposterior (\textit{y}) directions. The data are presented for: healthy senior controls (HSC), healthy young controls (HYC) and patients with Parkinson's disease (PDP) as mean (standard deviation). The values of parameters for open and closed eyes were used to calculate the corresponding Romberg ratio $R$ and stabilization ratio $SR$. }
   \label{tab:statistics}
\begin{tabular}{ccccccc}
   \toprule
   Group & $|v_x|$ [cm/s]  & $|v_y|$ [cm/s]  & $LZC_x$ & $LZC_y$
   & $BSV_x$ [cm/s] & $BSV_y$ [cm/s]
   \\

   \hline 
 &  &  &{\textbf{open eyes}} &  &   \tabularnewline

   HSC & 0.62 (0.2) & 0.57 (0.2) & 0.19 (0.02) & 0.20 (0.02) & 1.05 (0.4) & 1.05 (0.3)\\
   HYC  & 0.42 (0.2) & 0.41 (0.1)  & 0.19 (0.02) & 0.19  (0.02) & 0.76 (0.2) & 0.75 (0.2)\\
   PDP & 0.65 (0.5)  & 0.61 (0.3) & 0.16 (0.03) & 0.18 (0.02) & 1.20 (1.0) & 1.49 (1.2)\\
   
 &  &  &{\textbf{closed eyes}} &  &   \tabularnewline
   
   HSC & 0.89 (0.4) & 0.94 (0.3) & 0.21 (0.02) & 0.21 (0.02) & 1.64 (0.8) & 1.93 (1.0)\\
   HYC  & 0.64 (0.3) & 0.64 (0.2)  & 0.19 (0.02) & 0.19 (0.02) & 1.08 (0.4) & 1.11 (0.3)\\
   PDP & 0.80 (0.7)  & 0.89 (0.3) & 0.18 (0.02) & 0.19 (0.02) & 1.51 (1.3) & 2.15 (1.7) \\

 &  &  &{\textbf{Romberg ratio R}} &  &   \tabularnewline
   
   HSC & 1.43 (0.3) & 1.69 (0.4) & 1.10 (0.1) & 1.04 (0.1) & 1.54 (0.4) & 1.79 (0.6)\\
   HYC & 1.56 (0.5) & 1.59 (0.3) & 1.02 (0.1) & 1.04 (0.1) & 1.44 (0.5) & 1.50 (0.4)\\
   PDP & 1.22 (0.4) & 1.49 (0.4) & 1.10 (0.1) & 1.04 (0.1) & 1.26 (0.3) & 1.48 (0.4) \\
   
   &  &  &{\textbf{Stabilization ratio SR}} &  &   \tabularnewline
   
   HSC & 0.21 (0.2) & 0.31 (0.1) & 0.07 (0.1) & 0.03 (0.1) & 0.22 (0.1) & 0.29 (0.1)\\
   HYC & 0.25 (0.2) & 0.29 (0.1) & 0.01 (0.1) & 0.02 (0.1) & 0.18 (0.2) & 0.23 (0.2)\\
   PDP & 0.09 (0.2) & 0.24 (0.1) & 0.08 (0.1) & 0.02 (0.1) & 0.11 (0.2) & 0.20 (0.1) \\
   
   \tabularnewline
   
\hline

\end{tabular} 
\end{table}
 
\begin{figure}[ht]
\centering
\includegraphics[scale=0.55]{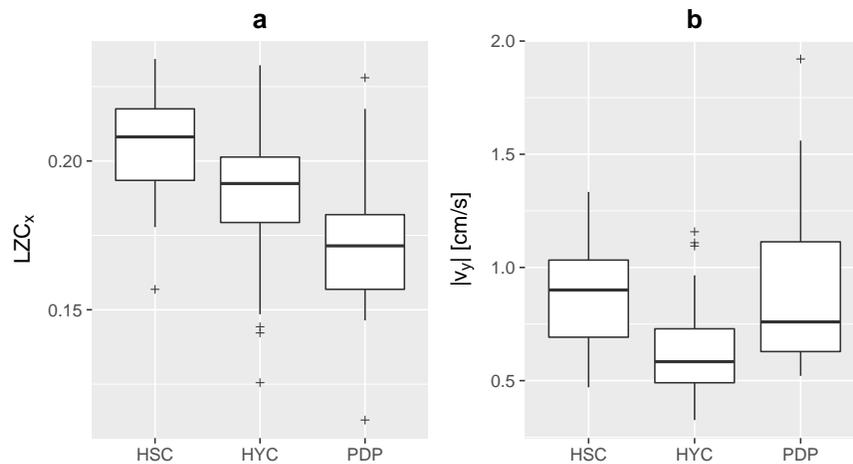}
\caption{Boxplots of: (a)  Lempel-Ziv complexity $LZC_x$ of mediolateral low-frequency velocity and (b) anteroposterior speed $|v_y|$. In both subplots data are presented for: healthy senior controls (HSC), healthy young controls (HYC) and patients with Parkinson's disease (PDP). }
\label{fig:LZC_V}
\end{figure}

For open eyes, mediolateral $LZC_x$ of elderly (HSC) and young (HYC) controls coincided: 0.19 (0.02) and were significantly higher ($p= 1 \times 10^{-6}$) than that of patients 0.16 (0.03). Posthoc analysis yielded $p= 7 \times 10^{-6}$ for comparison between HSC and PDP and   $p= 2 \times 10^{-5}$ for HYC vs PDP. Eye closure increased $LZC_x$ by 0.02 only in HSC and PDP, indicating the age-related influence of vision on  $LZC_x$. Lempel-Ziv complexity was the largest for HSC $LZC_x=0.21$ (0.02) and the lowest for PDP $LZC_x=0.18$ (0.02). The value for HYC fell in between $LZC_x=0.19$ (0.02). The  differences among the three cohorts were statistically significant ($p= 2 \times 10^{-7}$). The spread of $LZC_x$ in the eyes-closed condition is illustrated by boxplots in Fig. \ref{fig:LZC_V}a. $LZC_x$ differentiates three cohorts with only one exception: HYC vs HSC with eyes open (Table \ref{tab:significance}). The binary classification between HSC and PDP based on $LZC_x$ was most effective for closed eyes: sensitivity 92\%, specificity  87\%, F-score 89\% and area under curve (AUC) equal to 0.92. For open eyes, the discriminating power was also acceptable: sensitivity 81\%, specificity  83\%, F-score 81\% and AUC equal to  0.86.

We observed an age-related increase in both mediolateral and anteroposterior components of body sway velocity $BSV$ and speed $|v|$ (see Table \ref{tab:statistics}). For example, for closed eyes $BSV_y$ of healthy senior controls is 74\% higher than that of young adults (1.93 cm/s vs 1.11 cm/s). For the PDP cohort the difference is even stronger, being 94\%. The group-averaged low-frequency anteroposterior speed $|v_y|$ of elderly controls is 47\% higher that of young adults (0.94 cm/s vs 0.64 cm/s). The corresponding increase for PD patients is 39\% (Fig. \ref{fig:LZC_V}b). Table \ref{tab:significance} shows that age-related changes are statistically significant with  only two exceptions -- $BSV_x$ and $|v_x|$ for closed eyes.

Apart from $|v_x|$ in the eyes-closed condition, neither speed nor body sway velocity differentiate between PD patients and healthy senior controls. The 11\% decrease in $|v_x|$ for the PD cohort with respect to HSC is statistically significant but carries limited discriminating power. The binary classification based on $|v_x|$ has 61\% sensitivity, 76\% specificity, 69\% accuracy and AUC equal to 0.69.
 
For young adults, the absence of the influence of vision on complexity of low-frequency mediolateral velocity was reflected by the  group-averaged values of $SR_{LZC_x}=0.01 (0.1)$ and $R_{LZC_x}=1.02 (0.1)$. The influence of vision was stronger ($p= 7 \times 10^{-4}$ for three-way comparison) for both healthy senior controls $SR_{LZC_x}=0.07 (0.1)$ and PD patients $SR_{LZC_x}=0.08 (0.1)$ with no statistically significant difference between the latter groups. For mediolateral $BSV_x$ and anteroposterior $BSV_y$ sway velocities both the Romberg and stabilization ratios of the PD cohort were significantly smaller than those of healthy senior controls. For example, $SR_{BSV_x}$ for PDP 0.11 (0.2) was two times smaller than that of HSC 0.22 (0.1) with post hoc $p= 7 \times 10^{-3}$). The binary classification of HSC and PDP based on $SR_{BSV_x}$ had 69\% sensitivity, 80\% specificity, 75\% accuracy and AUC equal to 0.69.                            

\begin{table}
   \centering
   \caption{Statistical significance of differences in speed $|v|$, Lempel-Ziv complexity \textit{LZC} and body sway velocity $BSV$ in the open (OE) and closed (CE) eyes conditions. Subscripts \textit{x} and \textit{y} denote mediolateral and anteroposterior directions, respectively. A plus sign indicates that the \textit{p}-value from posthoc analysis was smaller than 0.05. An absence of statistically significant differences is indicated by a minus sign. The three cohorts are labeled as follows: HSC (healthy senior controls, HYC (healthy young controls), PDP ( patients with Parkinson's disease).} 
   \label{tab:significance}
\begin{tabular}{cccc}

   \toprule
   & HYC vs HSC  &  HYC vs PDP  & HSC vs PDP\\
   \hline 
   
  $|v_x|$ OE & + & + & -\\
  $|v_x|$ CE & + & - & +\\
  $|v_y|$ OE & + & + & -\\
  $|v_y|$ CE & + & + & -\\
  $LZC_x$ OE & - & + & +\\
  $LZC_x$ CE & + & + & +\\
  $LZC_y$ OE & + & - & +\\
  $LZC_y$ CE & - & - & -\\
  $BSV_x$ OE  & + & + & - \\ 
  $BSV_x$ CE & + & - & - \\
  $BSV_y$ OE & + & + & - \\ 
  $BSV_y$ CE  & + & + & - \\
   
\hline

\end{tabular} 
\end{table}

\section*{Discussion}

Age-related deterioration of postural stability is the major determinant of falls in older adults \cite{Melzer2004}. The significance of falls goes beyond high mortality rate and health care costs. The majority of older adults who have fallen experience psychological trauma which may cause fear of falling and the avoidance of physical activity, which in turn can result in muscle atrophy, sarcopenia and progressive substitution of muscle fibres by connective tissues. This vicious circle brings about both physical and psychological complications. During a 6-month follow-up,  50\% of PD patients had fallen and 25\% of them did so on multiple occasions. The corresponding rates for controls were 20\% for one-time fallers and 3\% for recurrent fallers. On an annual basis 70\% of PD patients fall and about half of them experience recurrent falls \cite{Bloem2004,pickering2007meta}. The mean age of onset of PD is around 60 years and the prevalence of disease increases with age. Thus, the age-related changes in balance control \cite{Goble2009,Deschamps2014,toledo2014age,Wiesmeier2015} form the backdrop for the analysis of pathological changes observed in PDP.

There are many measures of balance control derived from spontaneous fluctuations of the COP during quiet standing \cite{Maurer2005,kirchner2012evaluation}. In this study we investigated the properties of low-frequency ($<\sim1.5$ Hz) fluctuations of postural sway velocity. This choice was motivated by both theoretical and experimental evidence. In their seminal research, Collins and De Luca \cite{Collins1993,Collins1994} employed fractional Brownian motion to quantify inherent fluctuations of COP. They discovered the crossover point at $\sim1$s between the short-time persistent fluctuations with Hurst exponent $H > 0.5$ and long-term antipersistent fluctuations with $H < 0.5$.  They argued that balance regulation during quiet standing involves both open- and closed-loop control, the latter active over long-time intervals ($>1$s).  Delignieres et al. \cite{Delignieres2011} drew attention to both theoretical \cite{Kiemel2002} and experimental  \cite{Jeka2004} evidence that balance control is more likely to be velocity-based than position based. They developed a simple model for COP velocity dynamics, based on a bounded correlated random walk. This model reproduces a crossover effect, which for velocity time series is much more strongly pronounced than that of COP displacement time series. Peterka and Loughlin demonstrated that a relatively simple feedback model with a 150-200 ms time delay could account for postural control during spontaneous sway and a variety of perturbations \cite{Peterka2002,Peterka2004}. They argue that feedback is provided by a dynamically adjusted weighted combination of sensory orientation cues. This model suggests that insufficient sensorimotor regulation leads to resonant body sways with characteristic frequencies of 1 Hz. Masani et al. \cite{Masani2003} simultaneously measured EMG of ankle extensor, displacements of center of mass (COM) and COP during quiet standing. They concluded that a balance control strategy notably relies on velocity information and that muscle activity may be modulated in an anticipatory (feed-forward) manner. Maurer et al. found in the interval 0.7-1.1 Hz, peaks in the power spectrum of PD patients in the OFF state \cite{Maurer2003,Maurer2004}. The abnormally large and fast sways which contributed to such peaks vanished under levodopa treatment, deep brain stimulation (DBS) or a combination of both. The loss of oscillations and reduction of sway velocity were correlated with the improvement of patients' clinical motor assessment score. It is worth emphasizing that no such correlation was observed for sway amplitude. Moreover, patients reported improvements under therapy even though sway amplitude increased on average.

Body sway velocity (the ratio of the COP path length and the duration of the measurement) has long been recognized as a sensitive marker of developmental changes in body sway control \cite{Riach1987,Riach1994,Kirshenbaum2001,Rival2005,Chen2008,Ajrezo2013}. V\v sete\v ckov\'a and Frey recently measured body sway velocity in a cohort of 80 older adults, using the same narrow stance \cite{Vseteckova2013} as the one used in the present study. They found a statistically significant correlation between the age of subjects and both mediolateral and anteroposterior BSV. In the firm-surface, eyes-open trial, the values of Spearman-Brown rank correlation coefficients were equal to 0.686 and 0.727, respectively ($p < 0.001$). In the age interval of 70-74 years the median values of BSV were equal to 1.03 cm/s (mediolateral) and 1.01 cm/s (anteroposterior), in good agreement with the values obtained in our work for healthy senior controls (1.0 cm/s in both cases) whose mean age was 73 years. Please note that V\v sete\v ckov\'a and Frey reported only the median values.  

It is worth emphasizing that, with the exception of mediolateral speed $|v_x|$ in the eyes-closed condition, neither speed nor body sway velocity differentiate between PD patients and healthy senior controls. The 11\% decrease in $|v_x|$ for PD cohort with respect to HSC is statistically significant and is interesting in its own right. Melzer et al. \cite{Melzer2004} compared the quiet-stance posturographic measures of 19 elderly subjects who reported having fallen unexpectedly at least twice in the 6 months preceding the examination, with those of 124 non-fallers. In narrow base stance, the fallers had 28\% \textit{higher} mediolateral sway. The anteroposterior sway of both groups was not statistically different. It is also important to note that testing in wide stance was insufficient to discriminate elderly fallers. The BSV velocity (determined from the length of two-dimensional COP trajectory) in fallers was also higher (3.3 cm/s vs 2.6 cm/s, for a 20 s eyes-closed, firm-surface trial). In this study both  $|v_x|$ and $|BSV_x|$ were lower (albeit the 8\% decrease in the latter one was statistically insignificant) which is the first indication that impairment of balance control in PD patients has distinctive features not associated with aging.

The most important and novel finding of the present research concerns the influence of aging and Parkinson's disease on Lempel-Ziv complexity of low-frequency fluctuations of mediolateral velocity (Table \ref{tab:statistics}) $LZC_x$. For open eyes, $LZC_x$ of young and elderly controls 0.19 (0.02) was significantly higher than that of patients 0.16 (0.03). Eye closure increased $LZC_x$ by 0.02 only in HSC and PDP, indicating the influence of vision on  $LZC_x$  in elderly subjects. Such visual contribution is absent in young subjects. This interpretation is supported by the group averaged values of both the Romberg ratio $R$ and stabilization index $SR$. In particular, $SR$ of young adults is close to zero 0.01 (0.1). To put this value in perspective, this ratio for mediolateral speed $|v_x|$ is equal to 0.25 (0.2). The influence of vision on  $LZC_x$ in the elderly is reflected by the significantly higher $SR$ for HSC 0.07 (0.1) and PDP 0.08 (0.1) with no significant differences between the latter groups. In the eyes-closed condition the average value of $LZC_x$ for HSC is 11\% higher than that of HYC. For PDP we observed a 7\% decrease in  $LZC_x$  with respect to HYC which gives 18\% difference between PDP and HSC (Fig. \ref{fig:LZC_V}). One can see in Table \ref{tab:significance} that all three cohorts are statistically different. Thus, aging and PD have \textit{opposite} effects on the complexity of mediolateral velocity $v_x$. Moreover, this quantity has a good predictive power. It is possible to discriminate between PD patients and elderly controls with 77\% sensitivity, 93\% specificity, 86\% accuracy and AUC=0.87. The difference in  $LZC_x$ between PDP and HSC persists for open eyes. In this condition,  $LZC_x$  for PDP takes on the lowest value 0.16 (0.03). The predictive power is also good (although worse than in the closed-eyes condition) with 84\% sensitivity, 80\% specificity, 82\% accuracy and AUC=0.84. Averaging of $LZC_x$ for open and closed eyes leads to a well balanced classifier: 88\% sensitivity, 83\% specificity, 85\% accuracy and AUC=0.88.

The question arises as to the nature of the diminished mediolateral complexity in Parkinson's disease.  In the supplementary experiment, we measured body sway for 20 young controls standing on firm surface and foam which mimics impaired somatosensation. For closed eyes, the group averaged $LZC_x$ for the foam trial was 9.1\% lower than that for the firm surface with $p=0.004$ (a detailed discussion of the results of this study will be presented elsewhere). In principle, the reduced somatosensation could account for the 7\% difference between the HYC and PDP. It is worth noting that Manor et al. observed 9.4\% decrease of the value of complexity index of COP displacement, derived from multi-scale entropy (MSE), in the subjects with somatosensory impairment with respect to controls \cite{Manor2010}. Gow et al. reviewed applications of MSE to characterization of posturographic time series \cite{Gow2015}.

Priplata et al. \cite{Priplata2002,Priplata2003} demonstrated that the application of subsensory vibratory white noise can improve postural stability, possibly via a stochastic-resonance \cite{Collins1995}. In the search for a measure which would guide this novel therapeutic intervention, Costa et al. used MSE to characterize complexity of both COP displacement and velocity time series \cite{Costa2007}. They found that noise significantly increased the values of complexity across all MSE scales (ranging from 3 Hz to 100 Hz) both for young and elderly subjects. Please note that herein the upper edge of the frequency bandwidth was equal to 2 Hz. In the large longitudinal study Zhou et al. proved that older adults with lower baseline COP displacement complexity had greater probability of falling and that  traditional postural sway metrics did not associate with future falls \cite{Zhou2017}.

Deficits in balance and gait are common and disabling features of PD that significantly increase an individual’s risk of falling. There are a number of Parkinson's disease disability rating scales, nine of them are considered as recommended \cite{Shulman2016}.  While they are well suited to the clinical setting, by themselves, they have poor sensitivity and specificity for identifying prospective fallers in the PD population \cite{Kerr2010}. Using long measurement times and narrow stance, we found that with respect to healthy senior controls  patients with Parkinson's  disease had the diminished complexity of low-frequency mediolateral sway velocity and the lower average mediolateral speed. Thus, the reasons for the falls in PDP may well be very different from those associated with aging. This work contributes to the ongoing search for qualitative measures which are capable of tracking PD symptom progression. Further research is needed to establish whether a reduction of mediolateral complexity of low-frequency velocity reflects a propensity for falling and whether it is associated with the cognitive decline observed in many PD patients \cite{Deschamps2014, Kelly2015}.    

\section*{Methods}

\subsection*{Subjects}
The study was performed according to the Declaration of Helsinki and the protocol was approved by the Ethics Committee of Wroclaw Medical University. All subjects signed an informed consent. Neurologists specialized in extrapiramidal disorders diagnosed 30 PD patients (15 men, 15 women) according to the 2015 MDS clinical criteria and evaluated their clinical condition using the Unified Parkinson's Disease Rating Scale (MDS-UPDRS 2008), the Schwab and England Activities of Daily Living Scale (SEADLS) and the Hoehn-Yahr Standing Scale (H-YSS). The group average duration of PD was 9 (6) years (range 1-30 years) and the mean H-YSS stage was 2.5 (0.6). The average value of SEADLS was equal to 79 (8)\%. Part III of MDS-UPDRS varied from 15 to 52 with a mean value equal to 31 (12). 29 patients were treated with L-Dopa with an average daily dose of 523 (312) mg. 11 patients were given ropinirol, one of them in monotherapy. Piribedil was used in 2 patients, amantadine sulfate in 5 patients and biperiden in 2. 
The characteristics of PDP and HSC were as follows. PDP: age 73 (7) years, height 1.67 (0.1) m, weight 72 (11) kg, BMI 26 (3); HSC: age 71 (5) years, height 1.68 (0.09) m, weight 77 (12) kg, BMI 27 (4).
The HSC cohort was predominantly made up of the spouses of patients as well as members of the Seniors Club of the Wroclaw University of Science and Technology. Biomedical engineering students of this university participated in the study as healthy young controls: age 22 (2) years, height 1.73 (0.1) m, weight 71 (15) kg, BMI 23 (4). All subjects were screened to exclude those with a history of orthopedic problems, recent lower extremity injuries and those taking medications that could affect balance control.

\subsection*{Experimental task}
Participants were asked to wear comfortable shoes. They stood on a Wii Balance Board (WBB) in a narrow stance (heels and toes touching) with their arms resting alongside the body. This choice of stance was motivated by the outcome of the study of Melzer et al. of postural stability in the elderly \cite{Melzer2004}. The spontaneous fluctuations of center of pressure (COP) were recorded for 120 s in the eyes-open (OE) and eyes-closed (CE) conditions and were uniformly resampled at 50 Hz. There was a break between trials to avoid fatigue and boredom. During the first part the subjects looked at a fixation point located at a distance of 1.5 m, at individual eye level. The PD patients were tested in the morning during the ON phase of their pharmacotheraphy cycle (2-3 hours after taking medication). The applicability of the WBB to assessment of balance control was thoroughly proven \cite{Clark2010,Goble2014,Holmes2013}. In a recent review  of monitoring technologies used for the characterization of Parkinson's disease WBB was classified as recommended \cite{Godinho2016}.

\subsection*{Data analysis}

\begin{figure}[ht]
\centering
\includegraphics[scale=0.7]{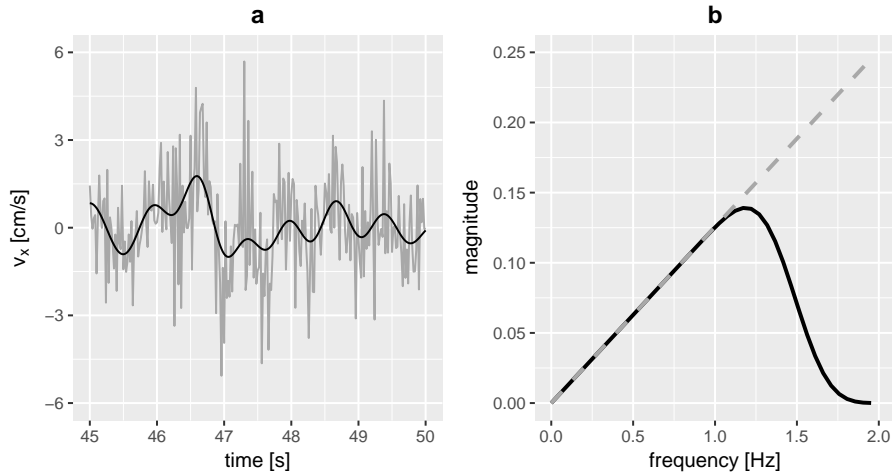}
\caption{(a) Mediolateral sway velocity calculated using a first-order Euler algorithm (gray line) and via differentiation with a low-pass differentiator filter (black line). (b) Zero-phase response of a FIR differentiator filter which was used to calculate mediolateral and anteroposterior velocities. The straight dashed line corresponds to the magnitude response of an ideal first-order differentiator.}
\label{fig:FirDif}
\end{figure}

Numerical differentiation of any physiological time series invariably leads to unwanted high frequency noise. This effect is elucidated by  Fig. \ref{fig:FirDif}a where  we compare  mediolateral sway velocity calculated using a first-order Euler algorithm (gray line) and via differentiation with a low-pass differentiator filter (black line). We used a FIR differentiator filter of order 300 with 2 Hz cut-off frequency and 0.8 Hz passband edge designed using a MATLAB R2016b designfilt function (equiripple algorithm with both passband and stopband weights equal to 1). The magnitude response of the filter is shown in Fig. \ref{fig:FirDif}b with a black line while a straight dashed line corresponds to the $i\omega$ frequency response function of an ideal first-order differentiator ($\omega$ being the angular frequency). The choice of filter parameters is to some extent arbitrary. We heuristically selected values that lead to the best discrimination between PDP and HSC. Wavelet differentiator filters are an alternative to FIR filters used in this study \cite{Gentile2003,Messina2004}. We found that the described FIR filter can be replaced by the one based on a continuous wavelet transform with a Gaussian mother function and scale parameter $a=10$ (for a sampling frequency 50 Hz) without affecting the values of the parameters analyzed in this study.

Using a FIR differentiator filter we calculated time series of low-frequency sway velocities: $ v_x$ (mediolateral direction) and $v_y$  (anteroposterior direction). For each direction, we computed the mean value of speed (absolute value of the velocity) and the value of the Lempel-Ziv complexity. To enable the comparison with previously published data we also calculated body sway velocity (BSV) defined as the length of the COP path (low-passed with fourth order Butterworth low-pass filter with cut-off frequency of 10 Hz) divided by the trial duration. The properties of Lempel-Ziv complexity are explained in detail in the last subsection.

Let $\mu$ be a posturographic measure which can be determined in the OE and CE conditions. The contribution of the visual system to balance maintenance is routinely estimated by the Romberg ratio (quotient):
\begin{equation}
R=\frac{\mu(CE)}{\mu(OE)}.
\label{R}
\end{equation}
Cornilleau-P\'er\'es et al. advocated the stabilization ratio SR
\begin{equation}
SR=1-\frac{\log(\mu(CE)+1)}{\log(\mu(OE)+1)}
\label{SR}
\end{equation}
as a more reliable estimator \cite{Cornilleau-Peres2005}. We use both $R$ and $SR$ to assess the role of visual system.

After performing the Shapiro-Wilk and Levene's tests, we chose Kruskal–Wallis test with Tukey’s post hoc comparisons to detect differences among cohorts. The significance threshold was set to 0.05.

\subsection*{Lempel-Ziv complexity}

Lempel and Ziv proposed a measure of algorithmic complexity which may be used to quantify the randomness of finite sequences \cite{Lempel1976}. This measure, which now bears their names, found numerous applications ranging from coding and data compression to biomedical signal processing \cite{Aboy2006}.  Lempel-Ziv Complexity (LZC)  is determined by the number of distinct substrings (words). For physiological time series, calculation of LZC is preceded by a coarse-graining. Typically, a signal of length $n$ is converted into a string of zeros and ones by comparison with the chosen threshold (e.g. the median) $S=s_1s_2 ... s_n$. We scan such a binary sequence $S$ from left to right in search of distinct words. Herein we use the variant of LZC algorithm known as LZ76 which decomposes $S$ into an exhaustive production process. Exhaustive complexity can be considered a lower limit of the complexity. The other production process discussed in the original paper -- primitive production process leads to the upper bound estimate.

\begin{figure}[ht]
\centering
\includegraphics[scale=0.45]{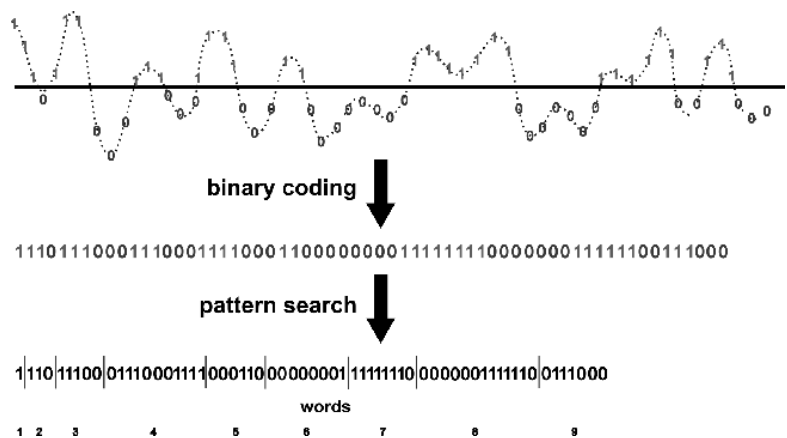}
\caption{The Lempel-Ziv complexity of a physiological time series is determined by a number of words found in a binary sequence obtained by comparison with the median. In the example there are 9 words in a sequence of length 62 so that $LZC=0.8643$}
\label{fig:LZC_diagram }
\end{figure}

We define LZC recursively. Let us assume that we scanned $r$ elements and we found $k$ words ${w}_{i}$. Said differently, $S(r)=s_1s_2... s_r$ is the concatenation of words $w_1w_2...w_k$. We then assign $Q=s(r+1)$ and check whether $Q$ is a substring of $SQ\pi$, where $\pi$ denotes the operation of deleting the last character in a sequence. $SQ$ is just the concatenation of $S$ and $Q$. If  $Q$ is a substring of $SQ\pi$ then we append $s(r+2)$ to $Q$: $Q=s(r+1)s(r+2)$ leaving $S(r)$ intact and check again whether $Q$ is a substring of $SQ\pi$. We repeat such process until a new word is found (which means that $Q$ is not a substring of $SQ\pi$) or the end of the sequence is reached. In the first case  $Q$ becomes a new word $w_{k+1}=Q$ and we append $Q$ to $S(r)$. In the latter case $Q$ becomes the last word. The first element of series $S(1)$ is always the first word. Fig. \ref{fig:LZC_diagram } elucides LZ76 algorithm. If $c$ is a number of words then $LZC$ of the sequence of length $n$ is defined as:
\begin{equation}
LZC=\frac{c \log_2(n)}{n}.
\label{LZC}
\end{equation}
In our calculations we used the implementation of the LZ76 algorithm proposed by Kaspar and Schuster \cite{PhysRevA.36.842} (please note that there is a typo in the algorithm's flowchart).
Aboy et al. \cite{Aboy2006} reviewed the application of LZC to biomedical time series analysis and through several numerical simulations demonstrated that it is particularly useful for estimating the bandwidth of a random process and the harmonic variability in quasi-periodic signals. LZC takes on the value of 0 for constant signal and 1 for noise.

\pagebreak
\bibliography{PD}

\section*{Acknowledgements}
Artur Maciejewski wrote a Matlab function which calculates LZC. Our undergraduate students Maciej Pilat, Monika Rachwalska, Anna Szczypka and Malgorzata Rakicka participated in posturographic measurements.

\section*{Author contributions statement}

M.L., S.B. and M.K conceived the experiment.  K.K., S.B. and M.K. conducted the experiment. M.L. K.K., and B.J.W. analysed the results.  All authors contributed to the preparation of the manuscript. 

\section*{Additional information}

\textbf{Competing financial interests} There were no conflicts of interest.

\end{document}